\documentclass[usenatbib]{mn2e}
\usepackage{mn2e_journal_style}
\usepackage{amssymb}
\usepackage{graphicx}

\newlength{\figurewidth}
\date{accepted 22 April 2009 - The definitive version is available at
       www.blackwell-synergy.com.}

\title[Gas accretion by massive star clusters]{Recurrent gas accretion 
  by massive star clusters, multiple stellar 
  populations and mass thresholds for spherical stellar systems}
\author[J.~Pflamm-Altenburg]
{Jan~Pflamm-Altenburg$^1$$^2$\thanks{email: jpflamm@astro.uni-bonn.de,
    pavel@astro.uni-bonn.de}
  and Pavel~Kroupa$^1$$^2$\footnotemark[1]\\
$^1$ Argelander-Institut f\"ur Astronomie, Universit\"at Bonn, 
  Auf dem H\"ugel 71, D-53121 Bonn, Germany\\
  $^2$ Rhine Stellar Dynamics Network (RSDN)
}
\begin{document}
\maketitle
\begin{abstract}
  We explore the gravitational influence of pressure supported
  stellar systems on the internal density distribution of a gaseous
  environment. We conclude that compact massive star clusters with 
  masses $\gtrsim 10^6\;M_\odot$ act as cloud condensation nuclei and
  are able to accrete gas recurrently  from a warm 
  interstellar medium which may cause further star formation events
  and account for multiple stellar populations in the most massive 
  globular and nuclear star clusters. The same analytical arguments 
  can be used to decide whether an arbitrary spherical stellar system is able
  to keep warm or hot interstellar material or not.
  These mass thresholds coincide with transition masses between pressure
  supported galaxies of different morphological types.
  
\end{abstract}
\section{Introduction}
Stars are believed to form during one single event 
in compact star clusters with a total mass
ranging from a few solar masses to a few million solar masses. 
But in recent years
observations have revealed that stellar populations of compact star clusters
are more complex than a single aged stellar ensemble. 

Very young star clusters such as the ONC \citep{palla2005a} and 
$\sigma$~Orionis \citep{sacco2007a} may probably harbour a few low mass
stars which are more than 10~Myr older than the main bulk of their stars. This
may be the result of extended star formation \citep{palla2005a} or the capture 
of stars, which are born in surrounding former star formation events,
through the deepening potential during cloud collapse
\citep{pflamm-altenburg2007a}. 

On the other hand some of the old massive globular clusters of both 
the Milky Way and Andromeda exhibit a spread in metallicity and/or 
have subpopulations with a helium overabundance suggesting
the occurrence of multiple star formation events in
these compact star clusters in the past (see Section~\ref{sec_evidence} for
a compilation of observational evidences).

All star clusters with evidences for multiple stellar
populations have in common that their total mass is around
10$^6$~$M_\odot$ or higher. Furthermore, star clusters with a mass larger
than $\sim$10$^6$~$M_\odot$ show a correlation between the cluster mass
and the cluster size, whereas star clusters less massive than 
$\sim$10$^6$~$M_\odot$ have a constant radius 
\citep{walcher2005a,hasegan2005a,dabringhausen2008a}. 

Mass thresholds of globular clusters have been considered
by several authors \citep{morgan1989a,shustov2000a,recchi2005a} 
addressing the question how massive a star cluster must be in order to
keep ejected material by AGB stars or supernovae.
However, these  thresholds can not explain the suggested growth of nuclear star
clusters by repeated accretion \citep{walcher2005a} and large internal spreads
in metallicity.

\citet{lin2007a} discussed the collective accretion of
gas by globular clusters onto their member stars embedded in a 10$^6$~K
hot back ground gas corresponding to todays circumstances. 
But globular clusters are assumed to have formed in denser environments,
as for example nuclear star clusters are forming today. 
Additionally, the treatment by \citet{lin2007a} does not reveal the 
existence of a mass threshold at 10$^6$~$M_\odot$.

In this paper we show that compact massive star clusters with masses 
$\gtrsim 10^6$~$M_\odot$ are able to accrete gas from an embedding warm
inter stellar medium.  
We first summarise the observational evidences for multiple stellar populations
in massive star clusters in Section~\ref{sec_evidence}. The theoretical 
criterion for possible gas accretion by massive star clusters is derived
in Section~\ref{sec_accretion} and the $10^6$~$M_\odot$ accretion mass
threshold is obtained in Section~\ref{sec_warm_threshold}. Additional 
mass thresholds are found and identified with transitions 
in the radius-mass plane of spherical and pressure supported stellar systems
(Section~\ref{sec_other_thresholds}).

\section{Evidence for multiple populations}
\label{sec_evidence}
The most massive globular cluster, $\omega$ Cen, with a mass of
$\sim$2.5$\cdot$10$^6$~$M_\odot$ \citep{mclaughlin2005a}  
shows a wide spread in metallicity \citep{freeman1975a}.
\citet{hilker2000a} determined by Str\"omgren photometry
three different stellar populations being $\sim$1--3~Gyr and $\sim$6~Gyr
younger than the oldest one. \citet{bedin2004a} confirmed the
existence of multiple evolutionary sequences in $\omega$ Cen.

The properties of RR~Lyrae stars observed in the globular clusters 
NGC~6388 and NGC~6441 with masses of $\sim$1.1$\times 10^6$~$M_\odot$
and  $\sim$1.6$\times 10^6$~$M_\odot$, respectively
\citep{mclaughlin2005a}, 
can be reproduced by the composition of two distinct 
populations \citep{ree2002a,yoon2008a}.  

Recently, \citet{milone2007a} showed that the split 
of the sub-giant branch of the 
Galactic globular cluster NGC~1851 with a mass of 
$\sim$3.1$\times 10^5$~$M_\odot$
\citep{mclaughlin2005a}
corresponds to the existence of
two different stellar populations with an age spread of
about 1~Gyr. Furthermore, \citet{milone2007a} concluded that 
the observed RR~Lyrea gap in NGC~1851 
requires an age difference of $\sim$2--3~Gyr.

\citet{lehnert1991a} reported a possible metallicity spread in M22
with a mass of 4.4$\cdot$10$^5$~$M_\odot$ \citep{mclaughlin2005a},
whereas \citet{richter1999a} found no evidence for a metallicity spread using
Str\"omgren photometry.

Metallicity spreads are also reported for G1 \citep{meylan2001a}, 
the most massive globular
cluster in M31 with a mass  of $\sim$8$\cdot$10$^6$~$M_\odot$ 
\citep{baumgardt2003b} and M54 \citep{sarajedini1995a}
with a mass of $\sim$2.0$\cdot$10$^6$~$M_\odot$ \citep{mclaughlin2005a}.

Evidence for metallicity spreads in the three massive globular clusters
G78, G213 and G280 of M31 are reported by \citet{fuentes-carrera2008a}.
These clusters have internal velocity dispersions as high as G1
\citep{djorgovski1997a} and therefore must have masses larger than
10$^6$~$M_\odot$. 

\begin{figure}
  \includegraphics[width=\columnwidth]{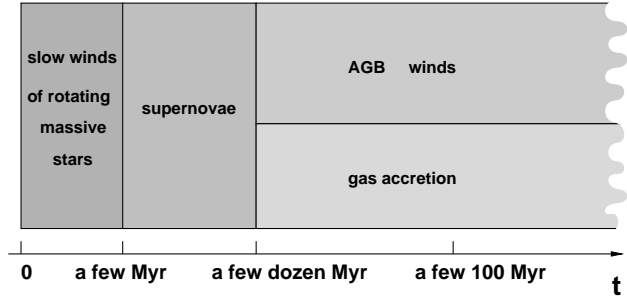}
  \caption{Sketch of different element and gas enrichment processes in
  massive star clusters.} 
  \label{fig_self_enrichment}
\end{figure}

Various explanations for the origin of multiple stellar populations in
globular clusters and abundance anomalies such as ejecta from AGB stars 
\citep[eg.][]{ventura2008a} or rotating massive 
stars \citep[eg.][]{decressin2007a} exist
(see \citealt{renzini2008a} for a summary). Self-enrichment by rotating 
massive stars works for the first few Myr (Fig.~\ref{fig_self_enrichment}) 
after star formation and might be attributed to observed 
anti-correlations of elements such as the Na-O
anti-correlation for star clusters with masses $ \gtrsim 10^4~M_\odot$.

Gas and metal return by supernovae occurs up to a few dozen Myr. As the 
ejecta by supernovae are much more energetic than winds by massive stars,
star cluster masses must be high in order to keep a significant fraction 
of the material supplied by supernovae. Using energy arguments, 
\citet*{baumgardt2008a} derived a lower mass threshold 
of $\approx 10^7~M_\odot$ for star clusters to retain
their residual gas despite multiple supernova events. \cite*{wuensch2008a} 
show using 
2D-hydrodynamical simulations that also lower mass star clusters may be able
to retain matter ejected by supernovae as a substantial fraction of 
such material can thermalise its high kinetic energy before escape 
from the star cluster.  

After all supernovae have exploded the massive AGB stars begin to evolve and 
are able to continuously replenish the gas reservoir in the star cluster and
in its vicinity \citep[e.g.][]{dantona2004a}. 

After the epoch of massive energy feedback by the radiation and winds of 
massive ionising stars and supernovae, which prevents gas of the 
surrounding ISM from being accreted, the gas of the ISM 
is able to react to the gravitational potential of the new star cluster.
It is expected that the more massive the star cluster is the stronger is the
gravitational influence on the ISM by the new star cluster. In this paper
we investigate this new possible scenario of gas accretion from the
surrounding ISM by massive star clusters.

Additionally, \citet{fellhauer2006a} showed 
that if $\omega$~Cen formed in a dwarf galaxy then it could have 
captured field stars during its formation from the underlying stellar 
field population contributing up to 40 per cent to  the total 
final star cluster 
mass and showing the complex stellar population composition of the 
host galaxy.  

All existing explanations have in common that the
material of further stellar generations has already assembled during
cluster formation and thus their total mass does not increase with time.

In contrast, \citet{walcher2005a} determined the masses of nuclear 
star clusters near the photometric centre of bulge-less spiral galaxies 
to lie between $0.8\times 10^6$~$M_\odot$ and
$6\times 10^7$~$M_\odot$. As these nuclear
clusters have luminosities by up to 2 orders of magnitude
larger than the most luminous Milky Way globular clusters,
\citet{walcher2005a} suggested that these nuclear clusters grow by 
repeated accretion of gas and show subsequent star formation.

\citet{boeker2004a} found that nuclear star clusters have dimensions
($r_\mathrm{hl}\approx 3.5$~pc)  
comparable to massive globular clusters, and some of these with masses 
$\gtrsim$10$^6$~$M_\odot$ show evidence for multiple 
structural components and estimated periods between star formation episodes
of about 0.5~Gyr \citep{seth2006a}.

\section{Model}
\label{sec_accretion}
From the hydrodynamical point of view the ISM
as a self-gravitating gas is described by the three conservation
equations for mass, momentum, and energy, the Poisson equation
and the equation of state \citep*[e.g.]{martel2006a},
\begin{equation}
  \frac{\partial\rho}{\partial t}+ \nabla\bullet
  \left(\rho\mathbf{v}\right) = 0\;,
\end{equation}
\begin{equation}
  \frac{\partial\mathbf{v}}{\partial t}+
  \left(\mathbf{v}\bullet\nabla\right)\mathbf{v}=
  -\frac{\nabla P}{\rho}-\nabla\Phi\;,
\end{equation}
\begin{equation}
  \frac{\partial \epsilon}{\partial t}+
  \mathbf{v}\bullet\nabla\epsilon = -P\nabla\bullet\mathbf{v}
  +\frac{\Gamma}{\rho}-\frac{\Lambda}{\rho}\;,
\end{equation}
\begin{equation}
  \nabla^2 \Phi = 4\pi G (\rho - \bar\rho)\;,
\end{equation}
\begin{equation}
  P = f(\rho,\epsilon)\;,
\end{equation}
where $\rho$ is the mass density, $P$ is the pressure, $\epsilon$ is the
specific energy, $\bf{v}$ is the velocity, $\Phi$ is the gravitational
potential, $\Gamma$ is the radiative heading rate and $\Lambda$ is the
cooling rate. $\bar \rho$ is the mean density. Following \citet{martel2006a}
this term has to be included to prevent the overall collapse of the
ISM and the term $-\bar\rho$ accounts for whichever process makes the ISM
globally stable. If a compact massive star cluster is present within
a homogeneous part of the ISM, then the potential in eq.~2 
has to be split into a self-gravitating part
and the external potential of the compact star cluster acting on the ISM.
If the density fluctuations $\rho - \bar\rho$ are initially small then 
the total potential in the vicinity of the star cluster is mainly given by
the star cluster itself.
To further simplify the problem we assume that the star cluster has no
relative velocity with respect to the embedding ISM, i.e. ${\bf v} = 0$. 
With increasing relative velocity between the star cluster and the embedding
ISM the effect of gas accretion  is expected to become less dominant.
Therefore, the stationary case gives a lower limit of a resulting mass
threshold.

As a final  simplification only the static case can be considered, i.e.
all time derivatives vanish. This can be interpreted such that the
out-of-equilibrium ISM, for which the density distribution does not 
follow the static solution, 
attempts to reach this state. Thus, the static solution can be used to
explore the gravitational influence of a compact massive star cluster
on the surrounding ISM and the full set of equations which  
are only numerically solvable reduces to the hydrostatic equation,
\begin{equation}
  \frac{1}{\rho}\nabla P = -\nabla \Phi_\mathrm{cl}\;, 
\end{equation}
\begin{equation}
  P = f(\rho,\epsilon)\;,
\end{equation}
where $\Phi_\mathrm{cl}$ is the potential of the massive star cluster.

The potential of the star cluster is assumed to be spherical. Then
for a simple isothermal ideal gas,
\begin{equation}
 P = \rho \frac{k_\mathrm{B}}{\mu m_\mathrm{u}} T\;,
\end{equation} 
with a temperature $T$ and a molecular mass  $\mu m_\mathrm{u}$  
of the isothermal ISM, the static solution is given by
\begin{equation}
  \label{eqn_solution}
  \rho(r) = \rho_0 \;e^{ 
    -\;\frac{\mu m_\mathrm{u}}{k_\mathrm{B} T}
    \;\left(\Phi_\mathrm{cl}(r)-\Phi_\mathrm{cl,0}\right)}\;,
\end{equation}
where $\rho_0$ and $\Phi_\mathrm{cl,0}$ are the mass density and 
the star-cluster potential at a given reference radius, $r_0$, introduced
by the integration. The integration constants in 
eq.~\ref{eqn_solution} are chosen such that the potential vanishes at infinity,
i.e. for a large distance from the cluster centre, and the particle density 
is the particle density of the undisturbed ISM ,
\begin{equation}
  \rho(\infty) = \mu m_\mathrm{u} n_\mathrm{ISM}\;\;,
  \;\;\Phi_\mathrm{cl,0} = \Phi(\infty)=0\;,
\end{equation}
where $n_\mathrm{ISM}$ is the particle density of the ISM.

The potential created by the compact star cluster is described
analytically by a Plummer potential \citep{plummer1911a},
\begin{equation}
  \label{eqn_plummer}
  \Phi_\mathrm{cl}(r) = - G\;M_\mathrm{cl}\;
  \left(r^2+b^2_\mathrm{cl}\right)^{-\frac{1}{2}}\;,
\end{equation} 
where $M_\mathrm{cl}$ is the mass of the star cluster and 
$b_\mathrm{cl}$ is the Plummer  parameter describing the compactness
of the cluster potential and is equal to the projected half-mass radius
\citep{heggie2003a}.

The final expression of the static solution is 
\begin{equation}
  \label{eq_final}
  n(r)=n_\mathrm{ISM}\;e^{\frac{G m_\mathrm{u}}{k_\mathrm{B}}\;\frac{\mu M_\mathrm{cl}}{T}\;
  \left(r^2+b_\mathrm{cl}^2\right)^{-\frac{1}{2}}}\;.
\end{equation}
The composition of the ISM is assumed to be primordial with $X=0.75$ 
relative mass fraction of hydrogen and $Y=0.25$ for helium. This corresponds
to a mean molecular mass of $\mu=4/(4X+Y)=1.23$. In the following the solutions for ISM-star-cluster systems
are plotted as a function of the star cluster mass and 
are parametrised by the ISM temperature and/or the Plummer parameter.
If the ISM is enriched to for example $Y=0.3$ then the mean molecular mass
is $\mu=1.29$, and the solutions correspond to the solution with the primordial
composition but a star cluster mass reduced by 4.8 per cent.
In the $n$-$M_\mathrm{cl}$ plot the curves then have to be shifted by 0.02 dex 
to the left. Thus the choice of a primordial ISM composition is appropriate 
for this analysis.

\section{The 10$^6$~M$_\odot$-mass threshold}
\label{sec_warm_threshold}
The Plummer  parameter of a star cluster is equal to the projected 
half-mass radius \citep{heggie2003a}. Assuming that the mass-to-light 
ratio of a massive
star cluster is independent of the radius then the projected half-mass radius
is equal to the half-light radius.  For Galactic globular clusters the
median  half-light radius is about 3.2~pc and ranges from 2~pc to 4~pc 
\citep{hasegan2005a,dabringhausen2008a}. 
The interstellar medium of the Galaxy consists of 
three main components \citep{mckee1995a,ferriere2001a,cox2005a}:
a molecular and atomic cold component below 100~K but occupying
only 1--2~per~cent of the interstellar volume, a warm atomic
and ionised component between $6\times 10^{3}$~K and 10$^4$~K with a 
particle density of $\sim$0.1--1.0~cm$^{-3}$ corresponding to the
ionisation of hydrogen, and a hot component with
temperatures with $\gtrsim 10^6$~K and a particle density less than  $10^{-2}$~cm$^{-3}$ fed by
supernovae.
The warm and the hot components have roughly the same volume filling factor 
of about 50~per~cent. 
So it is most likely that star clusters are embedded in the
warm or hot component of the ISM. 

\begin{figure}
 \includegraphics[width=\columnwidth]{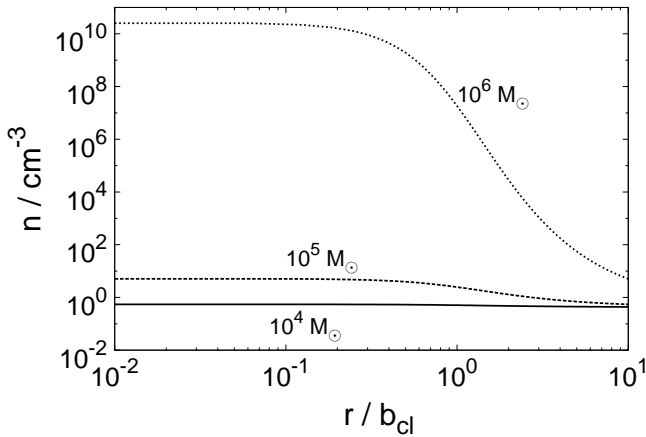}
  \caption{Radial density profile of the static solution 
    (eq.~\ref{eq_final}) for three different star cluster masses
    (10$^4$, 10$^5$, 10$^6$~$M_\odot$)
    and a Plummer parameter $b_\mathrm{cl}=3.2$~pc. The embedding ISM has a
    temperature of 8000~K and a particle density of 0.5~cm$^{-3}$.}
  \label{fig_radial_density}
\end{figure}
The radial gas density profiles of three different
star clusters ($M_\mathrm{cl}$ = 10$^4$, 10$^5$, 10$^6$~$M_\odot$) 
with a Plummer parameter $b_\mathrm{cl}$=3.2~pc embedded in a warm ISM 
with a temperature of 8000~K and a particle density
of 0.5~cm$^{-3}$ are plotted in Fig.~\ref{fig_radial_density}. It can be seen
that the hydrostatic gas density within the Plummer radius does not vary much.
We therefore use the central gas density, $n_\mathrm{c}=n(r=0)$, to characterise
the static gas density in the inner part of a star cluster. 

\begin{figure}
  \includegraphics[width=\columnwidth]{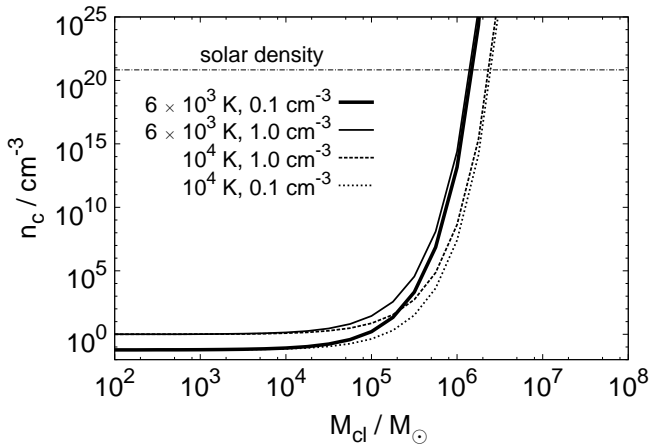}
  \caption{Central density in dependence of the star cluster mass with
    a Plummer radius of 3.2~pc for four different embedding interstellar
    media with a temperature of 6000 and 10$^4$~K and a density of
    0.1 and 1~cm$^{-3}$. The horizontal line denotes the average 
    density of matter in the solar interior.}
  \label{fig_central_density_b=3.2pc}
\end{figure}

In order to explore the influence of the cluster potential on the embedding
warm ISM we calculate the central density of the static solution 
(eq.~\ref{eq_final})
in dependence of the star cluster mass with a Plummer
radius of 3.2~pc shown in Fig~\ref{fig_central_density_b=3.2pc}.
To cover the full range of the observed warm interstellar medium
four different models are calculated with temperatures of 6000~K and
10$^4$~K and densities  of 0.1~cm$^{-3}$ and 1~cm$^{-3}$ corresponding
to the boundary values of the observed warm ISM. 
For cluster masses smaller than 10$^5$~$M_\odot$ the central particle
density does not differ from the density of the undisturbed ISM.
Between 10$^5$ and 10$^6$~$M_\odot$ the central density starts to rise
and becomes dramatically large above 10$^6$~$M_\odot$. 

The potential of massive star clusters with a mass larger than 10$^6$~$M_\odot$
thus produces an instability in the warm ISM such that the ISM is expected
to react with starting inflow towards the cluster centre. In other words:
A self-gravitating ISM does not notice the existence of star clusters with
a total mass lower than 10$^6$~$M_\odot$, but star clusters more massive than 
10$^6$~$M_\odot$ become immediately attracting holes in the warm ISM. 
Note, how rapidly the instability rises with increasing cluster mass 
only depends on the temperature but not on the density of the ISM. 
\begin{figure}
  \includegraphics[width=\columnwidth]{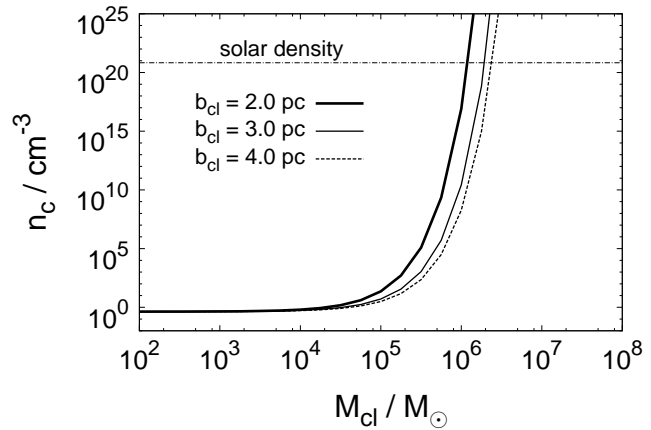}
  \caption{Central density in dependence of the star cluster mass for 
    three different Plummer radii of 2, 3.2 and 4~pc embedded in an 
    ISM with $T$ = 8000~K and $n$ = 0.5~cm$^{-3}$.}
  \label{fig_central_density_b_varied}
\end{figure}
Keeping the properties of the ISM constant ($T$ = 8000~K and 
$n$ = 0.5~cm$^{-3}$) the instability region also appears  
at cluster masses of about 10$^{6}$~$M_\odot$ when varying the 
Plummer parameter between 2 and 4~pc
(Fig.~\ref{fig_central_density_b_varied}).

Present-day Milky Way globular clusters do not belong to the disk and are not 
embedded in a comoving warm ISM. But in early times when they formed
they may have been embedded in a warm ISM, e.g. in a dwarf proto-galactic
building block. $\omega$~Cen for example is believed
to have been hosted in a dwarf galaxy disk before being accreted by the Milky
Way. Because it was massive enough it could have accreted additional gas when 
the conditions were appropriate. Star formation events within  
the young $\omega$~Cen or in its vicinity are expected to have caused
varying ISM properties. Thus it could have been placed alternatingly in 
the warm or hot ISM and the accretion history could have been fluctuating. 
Ionising OB stars in the young $\omega$~Cen would have prevented it from 
further accretion from the embedding ISM and star formation rested in 
$\omega$~Cen. Meanwhile star formation would have been ongoing 
in the ISM of the dwarf galaxy
disk being continuously enriched with metals. After all ionising sources
in $\omega$~Cen disappeared  accretion would have restarted if the 
coomoving conditions were appropriate. When enough gas accumulated in 
$\omega$~Cen star formation would have restarted and the newly
formed stars would have been more metal rich than the younger
stellar populations. Thus, distinct populations with different
metallicties are expected in such a scenario in $\omega$~Cen as observed.

\begin{figure}
 \includegraphics[width=\columnwidth]{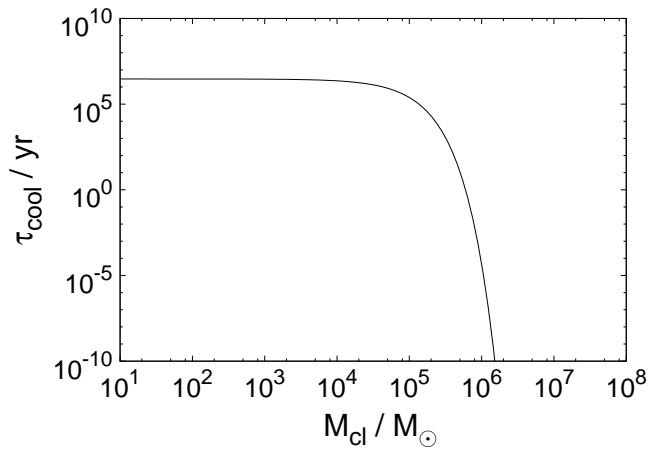}
  \caption{Cooling time-scale of the central static gas density in dependence
    on the star cluster mass for a Plummer parameter $b_\mathrm{cl}=3.2$~pc. 
    The embedding ISM has a temperature of 8000~K and a particle density of 
    0.5~cm$^{-3}$.}
  \label{fig_cool}
\end{figure}

To underline the strength of this instability threshold for star cluster
masses $\gtrsim 10^6 M_\odot$ the cooling time scale, $\tau_\mathrm{cool}$, 
corresponding to the
central gas density is calculated. We
use the cooling function provided in \citet*{koeppen1995a} in the temperature
regime from 100 to 10$^4$~K giving an estimate of the cooling time-scale
of 
\begin{equation}
\tau_\mathrm{cool} / \mathrm{yr} = 16400 \left(T/\mathrm{K}\right)^{0.5}
\left(n/\mathrm{cm}^{-3}\right)^{-1}\;.
\end{equation}
The resulting central  cooling time-scale for the static solution of an ISM with a temperature  of 8000~K and a density of 0.5~cm$^{-3}$ is plotted in
dependence of the star cluster mass with a Plummer parameter of 3.2~pc
(Fig.~\ref{fig_cool}). If the gas of the ISM starts to flow into the massive 
star cluster due to the instability caused by the cluster potential the
increasing central gas density has a decreasing cooling time-scale
supporting the gas accretion. For comparison, a thermal instability 
leading to an expected cooling flow of hot gas is known for
galaxy clusters \citep{fabian2003b} and massive elliptical galaxies
\citep{kroupa1994a}. New star formation is inhibited by the ISM in 
massive elliptical galaxies being kept at a temperature of 
$\approx 10^{6}$~K due
to the random stellar motions with velocities of a few hundred km~s$^{-1}$
\citep{mathews2003a,parriott2008a}. The massive-star-cluster instability, however, leads to a cooling
instability and the onset of star formation until the new OB~stars reheat
the cluster ISM.

The argument can also be turned around. If warm material
is released inside the cluster then the gas
attempts to reach the hydrostatic solution. The warm gas of a star
cluster less massive than 10$^6$~$M_\odot$ will be distributed nearly
uniformly in space, i.e. it escapes from the cluster, whereas a star
cluster more massive than 10$^6$~$M_\odot$ is able to keep its warm gas.

It may be rashly argued that the reason for accretion or
keeping of warm material is that the sound speed of the gas is smaller
than the escape velocity from the star cluster. The sound velocity of
an ideal gas is given by 
\begin{equation}
c_\mathrm{s} = \sqrt{\frac{\gamma k_\mathrm{B} T}{\mu m_\mathrm{u}}}
=91.2 \frac{\mathrm{m}}{\mathrm{s}}\sqrt{\frac{\gamma}{\mu}} 
\sqrt{\frac{T}{\mathrm{K}}}\;.
\end{equation} 
For a monatomic gas, $\gamma=5/3$, the sound speed is about 10.6~km/s
($T=10^4$~K) or 8.2~km/s ($T=6000$~K). The escape velocity at a distance
of $r=l b_\mathrm{cl}$ from the centre of a Plummer potential with 
mass $M_\mathrm{cl}$ and Plummer parameter $b_\mathrm{cl}$ is
\begin{equation}
  v_\mathrm{esc} = \sqrt{\frac{2G}{\sqrt{1+l}}
    \frac{M_\mathrm{cl}}{b_\mathrm{cl}}}
= 92.7 \frac{\mathrm{km}}{\mathrm{s}} \left(1+l\right)^{-\frac{1}{4}}
\sqrt{\frac{M_\mathrm{cl}}{M_\odot}\frac{\mathrm{pc}}{b_\mathrm{cl}}}\;.
\end{equation}
The lowest threshold mass of $2.5\times 10^4 M_\odot$
is calculated in the case of the lowest 
temperature of the warm ISM (6000~K) and the central escape velocity 
($l=0$). The largest 
threshold mass of $6.4\times 10^4 M_\odot$
is calculated in the case of the highest 
temperature of the warm ISM ($10^4$~K) and an escape
velocity from the outer region of the cluster, e.g. the half mass radius
($l=1.305$).  Nevertheless, by comparing these two 
characteristic velocities describing
the potential of the star cluster and the internal energy content of the gas
threshold masses are obtained which are more than one order of magnitude
smaller than the lower limit of masses of star clusters which multiple stellar 
population are observed in. In any case, the sound speed of a gas describes
the velocity with which a pressure change propagates through space. 
No large scale motion of
matter is involved in the propagation of sound, whereas the accretion of 
material is such a process. 

In any real situation of interest the gas will not be static 
and the interstellar medium will be turbulent and clumpy. Thus, it might
be questionable how much is learned by computing static gas densities.
Indeed, the calculated static solutions are not reached by the gas. 
The static solution describes a stage of an equilibrium. In the case of the
presence of a star cluster less massive than 10$^6 M_\odot$ this equilibrium 
is described by a nearly uniform density distribution. I.e. the warm ISM
does not notice the existence of the star cluster. In the case of a
star cluster more massive than 10$^6 M_\odot$ the static solution is 
charactarised by a required central gas density many orders of magnitude 
larger than the surrounding gas density. At the position of the massive 
star cluster a uniformly distributed ISM is far from equilibrium. Thus, the
gas tries to asymptotically reach this equilibrium by increasing its density
at the position of the massive star cluster, i.e. the star cluster accretes.
It will never reach the exact static gas density distribution as the increase
of the central gas density will result in a more efficient cooling and restarted
star formation will stop further accretion by heating the infalling gas.

Turbulences and inhomogeneities of the warm interstellar medium
lead to condensation of material and the formation of cold molecular clouds
at random locations in the galaxies. In this context massive compact star
clusters with  masses $\gtrsim 10^6 M_\odot$ stimulate this process at certain
locations, i.e. the star clusters act as cloud condensation
nuclei.

\section{The amount of accreted mass by a massive star cluster}
The threshold criterion derived here can be used to decide whether a star
cluster is able to accrete gas from a warm medium or not. If a massive 
star cluster is able to accrete the question arises how much material
can be accreted and if the amount of accreted material can account
for the observed multiple stellar populations. Finding an accurate answer
to this question requires extensive numerical simulation. However, a
reasonable estimation can be done using the accretion rate expression
for a star from its surrounding medium \citep{bondi1952a}. The star has
a potential of a point mass, whereas the potential of the star cluster
is extended. But due to the spherical symmetry of the star cluster,
the potential at a certain radius from the centre of the potential 
is given by the enclosed mass within this radius placed at 
the origin of the 
potential (Newtons's second theorem). 
Therefore, outside the cluster radius the cluster potential 
is the same as if the cluster is assumed to be a point mass and
the application of the Bondi-accretion is justified. The accretion rate, $A$,
is given by 
\begin{equation}
  A = 2 \pi \left(G M_\mathrm{cl}\right)^2 c_\mathrm{s} \rho_{\infty}\;,
\end{equation}  
where $c_\mathrm{s}$ is the sound speed and $\rho_{\infty}$ is the 
mass density of the gas far from the cluster, i.e. the density of the 
surrounding ISM.
With
\begin{equation}
\frac{A}{M_\odot / \mathrm{Myr}} = 2.9\times10^{-6} \mu 
\left(\frac{M_\mathrm{cl}}{M_\odot}\right)^2 
\left(\frac{c_\mathrm{s}}{\mathrm{km} / \mathrm{s}}\right)^{-3}
\frac{n_\mathrm{ISM}}{\mathrm{cm}^{-3}}
\end{equation}
and for a metallicity of $\mu=1.23$, a sound speed of 
$c_\mathrm{s} = 10.2$~km/s and a particle density of 
$n_\mathrm{ISM}=1$~cm$^{-3}$ the accretion rate is about 3400~$M_\odot$/Myr.
Thus on a time scale of a few 100~Myr corresponding to derived age
gaps between the multiple stellar populations the possible 
amount of accreted material is of the order of $3.4\times 10^5$~$M_\odot$ 
for a 10$^6$~$M_\odot$ star cluster. The accreted material can account
for the observed multiple stellar populations.

We note in passing that the accretion may also be modulated by the 
cluster orbit about its host galaxy. For example it may experience more 
significant accretion events when it passes through the ends of a 
galactic bar or through major spiral arms where the likelihood of it 
encountering cold gas is enhanced.

\section{On the cold-gas threshold for star clusters and 
the star-cluster birth instability}
The criterion above which star cluster mass a density instability of the
ISM can arise depends only on the ratio, $n_\mathrm{c}/n_\mathrm{ISM}$, of the
central and surrounding gas density in eq.~\ref{eq_final}. The threshold
expression is proportional to $e^{M_\mathrm{cl}/T b_\mathrm{cl}}$. Thus, it 
should be easier to accrete cold cloud material than warm  material. 
But the volume filling factor of cold cloud material in galaxies
is about 1--2 per cent and much smaller than the volume filling factor of
the warm material ($\approx$~50 per cent). The possibility that a massive 
star cluster is long-term embedded in a cold molecular cloud is much smaller
than the possibility that the star cluster is embedded in the warm ISM.
If a star cluster is surrounded by cold dense material it is expected that
accretion occurs for even less massive star clusters. 
As shown above the threshold is surpassed for typical values of 
$M_\mathrm{cl} = 10^6 M_\odot$, $b_\mathrm{cl}=3$~pc , and $T=10^4$~K. 
Star clusters should be able to accrete cold 
material with a temperature of about 100~K 
if the their mass-size ratio is larger than 
$M_\mathrm{cl}/b_\mathrm{cl} = 10^4~M_\odot/3~\mathrm{pc}$.  For young embedded
star clusters with $b_\mathrm{cl} = 0.3$~pc (as the ONC for example) a 
total cluster mass of about $1000 M_\odot$ is sufficient to accrete cold gas. 

It may be possible that this characterises the transition from distributed star 
formation within giant molecular clouds in the form of loose groups hosting
a few low-mass stars to star 
formation in well defined compact embedded star clusters: star formation 
in giant molecular clouds starts with distributed loose groups of a few 
low-mass stars. If regions in the low-mass-star forming giant molecular cloud
surpasse the threshold expression, run-away accretion starts and a well defined
star cluster appears. The ONC for example is a well defined young embedded star
cluster within a molecular cloud showing many loose groups of few low-mass stars
distributed over the molecular cloud.

\section{Nuclear star clusters}
Nuclear star clusters, which are embedded long-term  in a warm comoving ISM
and have masses less than $10^6$~$M_\odot$, may not accrete additional 
material after their formation. But nuclear star clusters which are born
or end up through mergers of smaller clusters  
more massive than $10^6$~$M_\odot$ should be able to accrete additional gas
after their formation and show further star formation events. Therefore
they are expected to grow as suggested by \cite{walcher2005a}. The
star cluster mass region at 10$^6$~$M_\odot$ should be underpopulated
predominantly for star clusters located in gas rich environments, i.e. the
central regions of galaxies. 

Indeed, the ensemble of
nuclear clusters observed by \citet{georgiev2009a} in gas rich dwarf 
irregulars have absolute V-band luminosities of $- 9.4$~mag and less 
corresponding to stellar masses of $\approx$~10$^6 M_\odot$ and more.  
Compared with the population of blue and red globular clusters hosted
in dIrrs, dEs and dSphs, which populate the distribution of V-band
luminosities uniformly down to $\approx -8.7$, 
these nuclear star clusters seem to form a distinct population as they are
separated from the rest of the less massive GCs
by a small gap in the luminosity sample \citep[][fig.~1]{georgiev2009a}.
These nucear star clusters  are candidates for 
repeated accretion events and so should contain complex stellar 
populations, as opposed to the clusters below the threshold which should 
not contain significant complex populations.  

It may be argued that nuclear star clusters are just fed with additional
gas because they are located at the origin of the potential of the 
galaxy. But the location of nuclear star clusters and the dynamical centers
of disk galaxies do not always coincide \citep{matthews2002a,walcher2005a}. 
Recently, \citet{georgiev2009b} found that seven of ten nuclear star clusters
in dwarf irregular galaxies are off-centre by up to 480~pc.  
The overall process can be as follows.

The central regions of galaxies are fed with gas due to the 
deep gravitational potential of the galaxy. For a nuclear star 
cluster to be able to accrete gas from this continuously refilled central
gas reservoir then depends only on the mass of the massive star cluster.
Due to differential rotation the accreted material is expected to form
a more rotationally than pressure supported new stellar system in the 
star cluster. Furthermore, younger stellar populations in these compact
massive star clusters should be more elliptical and aligned with the galactic
gas disk, whereas the initial population should be more 
spheroidal if the formation of the star clusters occurred in a monolithic 
collapse. Indeed it has been found that older stellar populations in nuclear
star clusters have a more spheroidal morphology and nuclear star clusters
appear flattened along the plane of the galaxy disk \citep{seth2008a}.

\citet{milosavljevi2004a} calculated the time-scale required for
nuclear star clusters to have migrated from distant eccentric orbits
in the disk towards the central regions of their host galaxies. Such 
migration time-scales are too long and nuclear star clusters must have 
formed in situ in the centres. However, if massive star clusters
accrete gas from the long-term embedding ISM their masses grew. 
If the angular momentum of the star cluster is conserved their orbit
must shrink and the inspiral may be accelerated. Places of the 
formation of such massive star cluster
could have been for example the massive clumps observed in chain galaxies at 
higher red shift \citep{elmegreen2006b}.

\section{Other mass thresholds}
\label{sec_other_thresholds}

In the previous sections the influence of a few pc sized star cluster
on the warm interstellar medium has been explored. But such globular/nuclear
star cluster type objects constitute only a subset of pressure
supported stellar systems, which do not have arbitrary dynamical properties
but show distinct size-mass relations \citep{hasegan2005a,dabringhausen2008a,forbes2008a}.

\begin{figure}
  \includegraphics[width=\columnwidth]{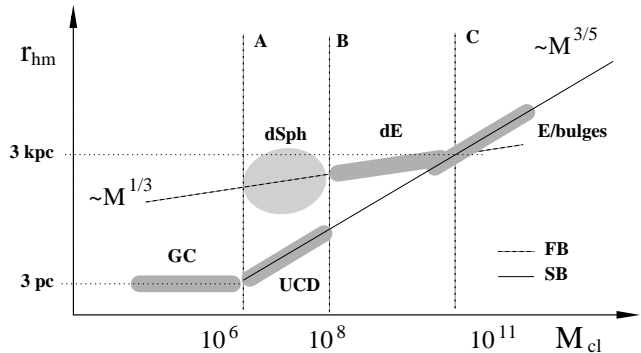}
  \caption{Illustration of the half-light radius dynamical
    mass relation of pressure supported stellar systems from 
    \citet{dabringhausen2008a}. See text for details.
  }
  \label{fig_dab_sketch}
\end{figure}

Figure~\ref{fig_dab_sketch} is a sketch of 
figure~2 of \citet{dabringhausen2008a} showing the 
half-light-radius--dynamical-mass relations of pressure supported 
stellar systems. Here, the dynamical mass, $M$, is the total mass within
the optical extend of the object as obtained by solving the Jeans equations.
Globular clusters do not show a mass-dependence of their half-light radius,
whereas above about $10^6$~$M_\odot$ the half-light radius becomes mass dependent.
The ultra compact dwarfs (UCDs), bulges of spiral galaxies 
and high-luminous ellipticals constitute a steep branch 
in the half-light--dynamical-mass plane. The radius-mass relation of these
objects lying on the steep branch (SB) presented in \citet{dabringhausen2008a} can be converted into a 
$b_\mathrm{cl}-M$ relation, 
\begin{equation}
\frac{b_\mathrm{cl}}{\mathrm{pc}} = 2.95 
\left(\frac{M}{10^6\;M_\odot}\right)^{0.60}\;.
\end{equation}

The flat branch (FB) is build by dwarf spheroidals and low-luminous ellipticals.
The  $b_\mathrm{cl}-M$ of the flat branch is
\begin{equation}
\frac{b_\mathrm{cl}}{\mathrm{pc}} = 92.8 
\left(\frac{M}{10^6\;M_\odot}\right)^{0.33}\;,
\end{equation}
corresponding to a constant mass density.

\begin{figure}
  \includegraphics[width=\columnwidth]{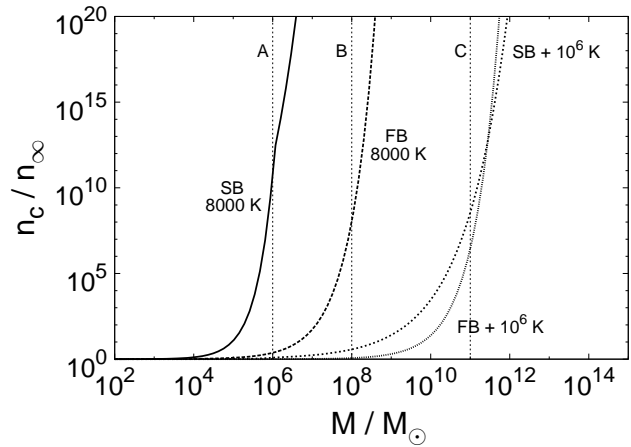}
  \caption{Density ratio of the static gas as a function of the total
    system mass of stellar systems lying on the flat (FB) and steep (SB)
    branch (Fig.~\ref{fig_dab_sketch}) for the cases of the warm and hot interstellar
    medium.
  }
  \label{fig_dab2}
\end{figure}

These two branches can be combined with the warm and the hot interstellar
medium. The ratio of the central density and the density of the gas 
at infinity of the static solution (eq.~\ref{eq_final}) 
is plotted in Fig.~\ref{fig_dab2} as a function of the
total mass of the system.  Systems 
with $M \ge 10^6$~$M_\odot$ on the
steep branch are able to keep their warm gas, whereas systems on the 
flat branch are able only to do so if their mass 
is $10^8$~$M_\odot$ or larger.
This mass threshold coincides with the mass separating the dwarf spheroidals
and low-luminous ellipticals from  each other. The hot interstellar medium 
can only be kept by systems more massive than about $10^{11}$~$M_\odot$.
This threshold mass corresponds to the transition region from low-luminous
ellipticals on the flat branch to high-luminous spirals and bulges of
disk galaxies on the steep branch.

One may speculate if these coincidences of the mass thresholds obtained 
from the instability argument and the mass transitions observed in the
mass-radius plane of pressure supported stellar systems has a direct physical
origin. 

\section{Conclusion}

We have shown that an ISM instability occurs for massive compact star clusters
with masses of $M_\mathrm{cl} \gtrsim 10^{6} M_\odot$. This instability may relate
to episodic gas accretion from the embedding ISM and subsequent star 
formation. Such an extended star formation history leads naturally to 
a spread in  metallicity as observed in globular star clusters
more massive than $\approx 10^6$~$M_\odot$.

This type of gas accretion can also
account for the ongoing star formation in massive nuclear star clusters
which are still embedded in a comoving dense interstellar medium.

The 'massive-star-cluster' instability adds to the well-known instabilities
of the ISM, e.g. the
Parker, Kelvin-Helmholz or Rayleigh-Taylor instability. 

Furthermore, for the combination of
observed size-mass relations of pressure supported systems and different
phases of the ISM  the same analysis reveals other mass thresholds which  
coincide with transitions in the size-mass diagram of pressure supported
stellar systems composed of globular clusters, ultra compact dwarf galaxies,
bulges of disk galaxies, dwarf spheroidals and dwarf and large ellipticals.

\vspace{0.5cm}
We thank Iskren Georgiev and J\"org Dabringhausen
for helpful discussion on nuclear and globular 
star clusters.
\bibliographystyle{mn2e}
\bibliography{sph,star-cluster,n-body,ism,onc,star-formation,r136,stellar-evolution,disk_galaxies,galaxy_clusters}

\begin{thebibliography}{}

\bibitem[\protect\citeauthoryear{{Baumgardt}, {Kroupa} \&
  {Parmentier}}{{Baumgardt} et~al.}{2008}]{baumgardt2008a}
{Baumgardt} H.,  {Kroupa} P.,    {Parmentier} G.,  2008, \mnras, 384, 1231

\bibitem[\protect\citeauthoryear{{Baumgardt}, {Makino}, {Hut}, {McMillan} \&
  {Portegies Zwart}}{{Baumgardt} et~al.}{2003}]{baumgardt2003b}
{Baumgardt} H.,  {Makino} J.,  {Hut} P.,  {McMillan} S.,    {Portegies Zwart}
  S.,  2003, \apjl, 589, L25

\bibitem[\protect\citeauthoryear{{Bedin}, {Piotto}, {Anderson}, {Cassisi},
  {King}, {Momany} \& {Carraro}}{{Bedin} et~al.}{2004}]{bedin2004a}
{Bedin} L.~R.,  {Piotto} G.,  {Anderson} J.,  {Cassisi} S.,  {King} I.~R.,
  {Momany} Y.,    {Carraro} G.,  2004, \apjl, 605, L125

\bibitem[\protect\citeauthoryear{{B{\"o}ker}, {Sarzi}, {McLaughlin}, {van der
  Marel}, {Rix}, {Ho} \& {Shields}}{{B{\"o}ker} et~al.}{2004}]{boeker2004a}
{B{\"o}ker} T.,  {Sarzi} M.,  {McLaughlin} D.~E.,  {van der Marel} R.~P.,
  {Rix} H.-W.,  {Ho} L.~C.,    {Shields} J.~C.,  2004, \aj, 127, 105

\bibitem[\protect\citeauthoryear{{Bondi}}{{Bondi}}{1952}]{bondi1952a}
{Bondi} H.,  1952, \mnras, 112, 195

\bibitem[\protect\citeauthoryear{{Cox}}{{Cox}}{2005}]{cox2005a}
{Cox} D.~P.,  2005, \araa, 43, 337

\bibitem[\protect\citeauthoryear{{Dabringhausen}, {Hilker} \&
  {Kroupa}}{{Dabringhausen} et~al.}{2008}]{dabringhausen2008a}
{Dabringhausen} J.,  {Hilker} M.,    {Kroupa} P.,  2008, \mnras, 386, 864

\bibitem[\protect\citeauthoryear{{D'Antona} \& {Caloi}}{{D'Antona} \&
  {Caloi}}{2004}]{dantona2004a}
{D'Antona} F.,  {Caloi} V.,  2004, \apj, 611, 871

\bibitem[\protect\citeauthoryear{{Decressin}, {Meynet}, {Charbonnel},
  {Prantzos} \& {Ekstr{\"o}m}}{{Decressin} et~al.}{2007}]{decressin2007a}
{Decressin} T.,  {Meynet} G.,  {Charbonnel} C.,  {Prantzos} N.,
  {Ekstr{\"o}m} S.,  2007, \aap, 464, 1029

\bibitem[\protect\citeauthoryear{{Djorgovski}, {Gal}, {McCarthy}, {Cohen}, {de
  Carvalho}, {Meylan}, {Bendinelli} \& {Parmeggiani}}{{Djorgovski}
  et~al.}{1997}]{djorgovski1997a}
{Djorgovski} S.~G.,  {Gal} R.~R.,  {McCarthy} J.~K.,  {Cohen} J.~G.,  {de
  Carvalho} R.~R.,  {Meylan} G.,  {Bendinelli} O.,    {Parmeggiani} G.,  1997,
  \apjl, 474, L19+

\bibitem[\protect\citeauthoryear{{Elmegreen} \& {Elmegreen}}{{Elmegreen} \&
  {Elmegreen}}{2006}]{elmegreen2006b}
{Elmegreen} D.~M.,  {Elmegreen} B.~G.,  2006, \apj, 651, 676

\bibitem[\protect\citeauthoryear{{Fabian}}{{Fabian}}{2003}]{fabian2003b}
{Fabian} A.~C.,  2003, \mnras, 344, L27

\bibitem[\protect\citeauthoryear{{Fellhauer}, {Kroupa} \& {Evans}}{{Fellhauer}
  et~al.}{2006}]{fellhauer2006a}
{Fellhauer} M.,  {Kroupa} P.,    {Evans} N.~W.,  2006, \mnras, 372, 338

\bibitem[\protect\citeauthoryear{{Ferri{\`e}re}}{{Ferri{\`e}re}}{2001}]{ferrie%
re2001a}
{Ferri{\`e}re} K.~M.,  2001, Reviews of Modern Physics, 73, 1031

\bibitem[\protect\citeauthoryear{{Forbes}, {Lasky}, {Graham} \&
  {Spitler}}{{Forbes} et~al.}{2008}]{forbes2008a}
{Forbes} D.~A.,  {Lasky} P.,  {Graham} A.~W.,    {Spitler} L.,  2008, \mnras,
  389, 1924

\bibitem[\protect\citeauthoryear{{Freeman} \& {Rodgers}}{{Freeman} \&
  {Rodgers}}{1975}]{freeman1975a}
{Freeman} K.~C.,  {Rodgers} A.~W.,  1975, \apjl, 201, L71+

\bibitem[\protect\citeauthoryear{{Fuentes-Carrera}, {Jablonka}, {Sarajedini},
  {Bridges}, {Djorgovski} \& {Meylan}}{{Fuentes-Carrera}
  et~al.}{2008}]{fuentes-carrera2008a}
{Fuentes-Carrera} I.,  {Jablonka} P.,  {Sarajedini} A.,  {Bridges} T.,
  {Djorgovski} G.,    {Meylan} G.,  2008, \aap, 483, 769

\bibitem[\protect\citeauthoryear{{Georgiev}, {Hilker}, {Puzia}, {Goudfrooij} \&
  H.}{{Georgiev} et~al.}{2009}]{georgiev2009b}
{Georgiev} I.~Y.,  {Hilker} M.,  {Puzia} T.~H.,  {Goudfrooij} P.,    H. B.,
  2009, \mnras, -- in press

\bibitem[\protect\citeauthoryear{{Georgiev}, {Puzia}, {Hilker} \&
  {Goudfrooij}}{{Georgiev} et~al.}{2009}]{georgiev2009a}
{Georgiev} I.~Y.,  {Puzia} T.~H.,  {Hilker} M.,    {Goudfrooij} P.,  2009,
  \mnras, 392, 879

\bibitem[\protect\citeauthoryear{{Ha{\c s}egan}, {Jord{\'a}n}, {C{\^o}t{\'e}},
  {Djorgovski}, {McLaughlin}, {Blakeslee}, {Mei}, {West}, {Peng}, {Ferrarese},
  {Milosavljevi{\'c}}, {Tonry} \& {Merritt}}{{Ha{\c s}egan}
  et~al.}{2005}]{hasegan2005a}
{Ha{\c s}egan} M.,  {Jord{\'a}n} A.,  {C{\^o}t{\'e}} P.,  {Djorgovski} S.~G.,
  {McLaughlin} D.~E.,  {Blakeslee} J.~P.,  {Mei} S.,  {West} M.~J.,  {Peng}
  E.~W.,  {Ferrarese} L.,  {Milosavljevi{\'c}} M.,  {Tonry} J.~L.,    {Merritt}
  D.,  2005, \apj, 627, 203

\bibitem[\protect\citeauthoryear{{Heggie} \& {Hut}}{{Heggie} \&
  {Hut}}{2003}]{heggie2003a}
{Heggie} D.,  {Hut} P.,  2003, {The Gravitational Million-Body Problem: A
  Multidisciplinary Approach to Star Cluster Dynamics}.
The Gravitational Million-Body Problem: A Multidisciplinary Approach to Star
  Cluster Dynamics, by Douglas Heggie and Piet Hut.~ Cambridge University
  Press, 2003, 372 pp.

\bibitem[\protect\citeauthoryear{{Hilker} \& {Richtler}}{{Hilker} \&
  {Richtler}}{2000}]{hilker2000a}
{Hilker} M.,  {Richtler} T.,  2000, \aap, 362, 895

\bibitem[\protect\citeauthoryear{{Koeppen}, {Theis} \& {Hensler}}{{Koeppen}
  et~al.}{1995}]{koeppen1995a}
{Koeppen} J.,  {Theis} C.,    {Hensler} G.,  1995, \aap, 296, 99

\bibitem[\protect\citeauthoryear{{Kroupa} \& {Gilmore}}{{Kroupa} \&
  {Gilmore}}{1994}]{kroupa1994a}
{Kroupa} P.,  {Gilmore} G.~F.,  1994, \mnras, 269, 655

\bibitem[\protect\citeauthoryear{{Lehnert}, {Bell} \& {Cohen}}{{Lehnert}
  et~al.}{1991}]{lehnert1991a}
{Lehnert} M.~D.,  {Bell} R.~A.,    {Cohen} J.~G.,  1991, \apj, 367, 514

\bibitem[\protect\citeauthoryear{{Lin} \& {Murray}}{{Lin} \&
  {Murray}}{2007}]{lin2007a}
{Lin} D.~N.~C.,  {Murray} S.~D.,  2007, \apj, 661, 779

\bibitem[\protect\citeauthoryear{{Martel}, {Evans} II \& {Shapiro}}{{Martel}
  et~al.}{2006}]{martel2006a}
{Martel} H.,  {Evans} II N.~J.,    {Shapiro} P.~R.,  2006, \apjs, 163, 122

\bibitem[\protect\citeauthoryear{{Mathews} \& {Brighenti}}{{Mathews} \&
  {Brighenti}}{2003}]{mathews2003a}
{Mathews} W.~G.,  {Brighenti} F.,  2003, \araa, 41, 191

\bibitem[\protect\citeauthoryear{{Matthews} \& {Gallagher} III}{{Matthews} \&
  {Gallagher}}{2002}]{matthews2002a}
{Matthews} L.~D.,  {Gallagher} III J.~S.,  2002, \apjs, 141, 429

\bibitem[\protect\citeauthoryear{{McKee}}{{McKee}}{1995}]{mckee1995a}
{McKee} C.~F.,  1995, in {Ferrara} A.,  {McKee} C.~F.,  {Heiles} C.,
  {Shapiro} P.~R.,  eds, The Physics of the Interstellar Medium and
  Intergalactic Medium Vol.~80 of Astronomical Society of the Pacific
  Conference Series, {The Multiphase Interstellar Medium}.
pp 292--+

\bibitem[\protect\citeauthoryear{{McLaughlin} \& {van der Marel}}{{McLaughlin}
  \& {van der Marel}}{2005}]{mclaughlin2005a}
{McLaughlin} D.~E.,  {van der Marel} R.~P.,  2005, \apjs, 161, 304

\bibitem[\protect\citeauthoryear{{Meylan}, {Sarajedini}, {Jablonka},
  {Djorgovski}, {Bridges} \& {Rich}}{{Meylan} et~al.}{2001}]{meylan2001a}
{Meylan} G.,  {Sarajedini} A.,  {Jablonka} P.,  {Djorgovski} S.~G.,  {Bridges}
  T.,    {Rich} R.~M.,  2001, \aj, 122, 830

\bibitem[\protect\citeauthoryear{{Milone}, {Bedin}, {Piotto}, {Anderson},
  {King}, {Sarajedini}, {Dotter}, {Chaboyer}, {Marin-Franch}, {Majewski},
  {Aparicio}, {Hempel}, {Paust}, {Reid}, {Rosenberg} \& {Siegel}}{{Milone}
  et~al.}{2007}]{milone2007a}
{Milone} A.~P.,  {Bedin} L.~R.,  {Piotto} G.,  {Anderson} J.,  {King} I.~R.,
  {Sarajedini} A.,  {Dotter} A.,  {Chaboyer} B.,  {Marin-Franch} A.,
  {Majewski} S.,  {Aparicio} A.,  {Hempel} M.,  {Paust} N.~E.~Q.,  {Reid}
  I.~N.,  {Rosenberg} A.,    {Siegel} M.,  2007, ArXiv e-prints, 709

\bibitem[\protect\citeauthoryear{{Milosavljevi{\'c}}}{{Milosavljevi{\'c}}}{200%
4}]{milosavljevi2004a}
{Milosavljevi{\'c}} M.,  2004, \apjl, 605, L13

\bibitem[\protect\citeauthoryear{{Morgan} \& {Lake}}{{Morgan} \&
  {Lake}}{1989}]{morgan1989a}
{Morgan} S.,  {Lake} G.,  1989, \apj, 339, 171

\bibitem[\protect\citeauthoryear{{Palla}, {Randich}, {Flaccomio} \&
  {Pallavicini}}{{Palla} et~al.}{2005}]{palla2005a}
{Palla} F.,  {Randich} S.,  {Flaccomio} E.,    {Pallavicini} R.,  2005, \apjl,
  626, L49

\bibitem[\protect\citeauthoryear{{Parriott} \& {Bregman}}{{Parriott} \&
  {Bregman}}{2008}]{parriott2008a}
{Parriott} J.~R.,  {Bregman} J.~N.,  2008, \apj, 681, 1215

\bibitem[\protect\citeauthoryear{{Pflamm-Altenburg} \&
  {Kroupa}}{{Pflamm-Altenburg} \& {Kroupa}}{2007}]{pflamm-altenburg2007a}
{Pflamm-Altenburg} J.,  {Kroupa} P.,  2007, \mnras, 375, 855

\bibitem[\protect\citeauthoryear{{Plummer}}{{Plummer}}{1911}]{plummer1911a}
{Plummer} H.~C.,  1911, \mnras, 71, 460

\bibitem[\protect\citeauthoryear{{Recchi} \& {Danziger}}{{Recchi} \&
  {Danziger}}{2005}]{recchi2005a}
{Recchi} S.,  {Danziger} I.~J.,  2005, \aap, 436, 145

\bibitem[\protect\citeauthoryear{{Ree}, {Yoon}, {Rey} \& {Lee}}{{Ree}
  et~al.}{2002}]{ree2002a}
{Ree} C.~H.,  {Yoon} S.-J.,  {Rey} S.-C.,    {Lee} Y.-W.,  2002, in {van
  Leeuwen} F.,  {Hughes} J.~D.,   {Piotto} G.,  eds, ASP Conf. Ser. 265: Omega
  Centauri, A Unique Window into Astrophysics {Synthetic Color-Magnitude
  Diagrams for {$\omega$} Centauri and Other Massive Globular Clusters with
  Multiple Populations}.
pp 101--+

\bibitem[\protect\citeauthoryear{{Renzini}}{{Renzini}}{2008}]{renzini2008a}
{Renzini} A.,  2008, accepted by \aap, 808

\bibitem[\protect\citeauthoryear{{Richter}, {Hilker} \& {Richtler}}{{Richter}
  et~al.}{1999}]{richter1999a}
{Richter} P.,  {Hilker} M.,    {Richtler} T.,  1999, \aap, 350, 476

\bibitem[\protect\citeauthoryear{{Sacco}, {Randich}, {Franciosini},
  {Pallavicini} \& {Palla}}{{Sacco} et~al.}{2007}]{sacco2007a}
{Sacco} G.~G.,  {Randich} S.,  {Franciosini} E.,  {Pallavicini} R.,    {Palla}
  F.,  2007, \aap, 462, L23

\bibitem[\protect\citeauthoryear{{Sarajedini} \& {Layden}}{{Sarajedini} \&
  {Layden}}{1995}]{sarajedini1995a}
{Sarajedini} A.,  {Layden} A.~C.,  1995, \aj, 109, 1086

\bibitem[\protect\citeauthoryear{{Seth}, {Blum}, {Bastian}, {Caldwell} \&
  {Debattista}}{{Seth} et~al.}{2008}]{seth2008a}
{Seth} A.~C.,  {Blum} R.~D.,  {Bastian} N.,  {Caldwell} N.,    {Debattista}
  V.~P.,  2008, \apj, 687, 997

\bibitem[\protect\citeauthoryear{{Seth}, {Dalcanton}, {Hodge} \&
  {Debattista}}{{Seth} et~al.}{2006}]{seth2006a}
{Seth} A.~C.,  {Dalcanton} J.~J.,  {Hodge} P.~W.,    {Debattista} V.~P.,  2006,
  \aj, 132, 2539

\bibitem[\protect\citeauthoryear{{Shustov} \& {Wiebe}}{{Shustov} \&
  {Wiebe}}{2000}]{shustov2000a}
{Shustov} B.~M.,  {Wiebe} D.~S.,  2000, \mnras, 319, 1047

\bibitem[\protect\citeauthoryear{{Ventura} \& {D'Antona}}{{Ventura} \&
  {D'Antona}}{2008}]{ventura2008a}
{Ventura} P.,  {D'Antona} F.,  2008, \aap, 479, 805

\bibitem[\protect\citeauthoryear{{Walcher}, {van der Marel}, {McLaughlin},
  {Rix}, {B{\"o}ker}, {H{\"a}ring}, {Ho}, {Sarzi} \& {Shields}}{{Walcher}
  et~al.}{2005}]{walcher2005a}
{Walcher} C.~J.,  {van der Marel} R.~P.,  {McLaughlin} D.,  {Rix} H.-W.,
  {B{\"o}ker} T.,  {H{\"a}ring} N.,  {Ho} L.~C.,  {Sarzi} M.,    {Shields}
  J.~C.,  2005, \apj, 618, 237

\bibitem[\protect\citeauthoryear{{W{\"u}nsch}, {Tenorio-Tagle}, {Palou{\v s}}
  \& {Silich}}{{W{\"u}nsch} et~al.}{2008}]{wuensch2008a}
{W{\"u}nsch} R.,  {Tenorio-Tagle} G.,  {Palou{\v s}} J.,    {Silich} S.,  2008,
  \apj, 683, 683

\bibitem[\protect\citeauthoryear{{Yoon}, {Joo}, {Ree}, {Han}, {Kim} \&
  {Lee}}{{Yoon} et~al.}{2008}]{yoon2008a}
{Yoon} S.-J.,  {Joo} S.-J.,  {Ree} C.~H.,  {Han} S.-I.,  {Kim} D.-G.,    {Lee}
  Y.-W.,  2008, \apj, 677, 1080

\end{thebibliography}
\appendix

\end{document}